# FPMax: a 106GFLOPS/W at 217GFLOPS/mm² Single-Precision FPU, and a 43.7GFLOPS/W at 74.6GFLOPS/mm² Double-Precision FPU, in 28nm UTBB FDSOI

Jing Pu[1], Sameh Galal[2], Xuan Yang[1], Ofer Shacham[3], Mark Horowitz[1]

[1]Stanford University, Stanford, CA USA, [2]Soft Machines Inc., Santa Clara, CA USA, [3]Google Inc., Mountain View, CA USA

## Abstract

FPMax implements four FPUs optimized for latency or throughput workloads in two precisions, fabricated in 28nm UTBB FDSOI. Each unit's parameters, e.g pipeline stages, booth encoding etc., were optimized to yield 1.42ns latency at 110GLOPS/W (SP) and 1.39ns latency at 36GFLOPS/W (DP). At 100% activity, body-bias control improves the energy efficiency by about 20%; at 10% activity this saving is almost 2x.

## Introduction

Floating-point (FP) computation has become ubiquitous in digital systems for either sequential or parallel workloads. To help designers create efficient implementations for these different environments, we created FPGen, an FPU generator, which extracted design innovations from 50 years of published research on FPU design [1]. FPGen explores different implementation techniques to find those that optimize the design for the desired applications' power, performance and area constraints. The FPMax chip evaluates the capability of FPGen and Ultra-Thin Body BOX fully-depleted SOI (UTBB FDSOI) technology by incorporating four generated FP Multiply-Accumulate (FMAC) units optimized for different precisions and either latency or throughput applications in ST 28nm UTBB FDSOI LVT technology.

All 4 FPUs are fully pipelined, implement IEEE compliant rounding, and utilize internal forwarding before rounding [8]. They use widely different implementations for the FMAC (Table I). The designs optimized to minimize the latency use a cascade multiply-add (CMA) architecture with a Wallace tree to sum up the partial products (PPs). The throughput optimized designs use a fused multiply accumulation (FMA) design with simpler combiners for the multiplication.

## FPU Architectures

For latency oriented FPUs, the primary metrics are the energy/FLOP and the average latency per FLOP. In FMAs [2], the latencies for using the result as a multiplier or an adder input are the same (Fig. 1(a)). However, in many applications, accumulation dependencies tend to be more common. A CMA architecture has a longer total latency, but a shorter path for accumulation (Fig. 1(b)), and thus performs better for practical workloads. Fig. 2(a) shows the pipeline of our 5-stage double-precision (DP) CMA. With internal bypasses, the un-rounded result at stage 4 can be forwarded either to the multiplier input at stage 1 or to the adder input at stage 3 or earlier. We define an average latency penalty as the average number of cycles a dependent operation (either accumulation or multiplication) must stall before its data is available [1]. Our experiments show that compared to 5-cycle FMAs with and without unrounded results forwarding, DP CMA achieves 37% and 57% less average latency penalty in SPEC FP benchmarks, respectively (Fig. 2(c)). For the single-precision (SP) latency optimized unit, we explored a more deeply pipelined and faster clocked design. The longer clock cycle allows the DP unit to use Booth 3 encoding to reduce the area and energy, while the SP unit uses more traditional Booth 2 encoding.

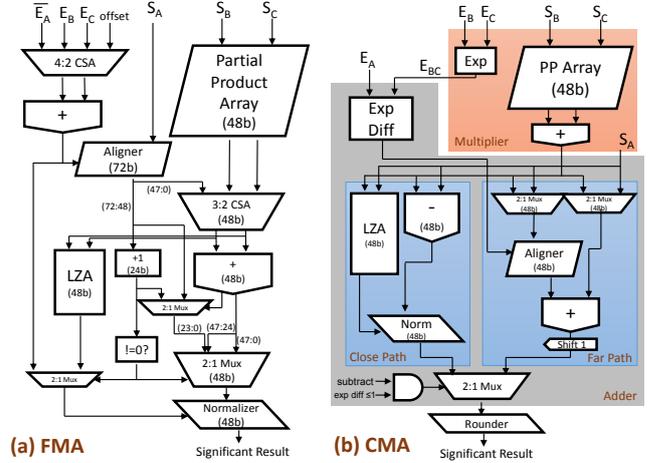

Fig. 1. Simplified block diagrams for single-precision (a) FMA and (b) CMA (adapted from [2]).

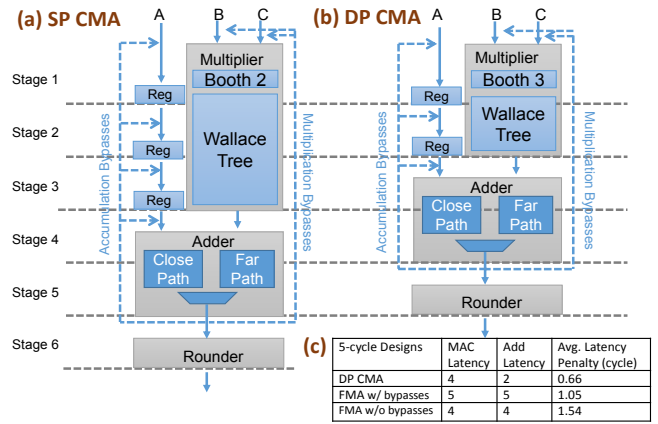

Fig. 2. (a) SP and (b) DP CMA pipelines and internal bypasses and (c) latency comparison for CMA and FMA w/ and w/o bypasses.

For GPU type applications with abundant parallelism, the latency of an individual operation is less critical. As a result, the metrics become the energy per FLOP and the compute efficiency in GFLOPS/mm² [2]. We find that FMAs are more area efficient than CMAs. The focus on area and energy efficiencies again leads to the use of Booth 3 encoding, and also simpler combiner structures for the multiplier partial products: the DP units uses a simple array and the SP uses a modified array called a ZM structure [3].

## Chip Implementation and Measurement

The design parameters of the FPUs were selected from the Pareto curves of energy vs. performance shown in Fig. 3 for SP throughput designs. The curve with triangle marks represents the performance for designs with different architectural parameters simulated at 1V supply using FPGen. Given the ability to change $V_{DD}$, the fabricated SP FMA performance is illustrated by the curve marked with white squares. In addition, adding the body-bias (BB) control of the UTBB FDSOI process improves energy efficiency (at a constant area) by 21%, or improves area efficiency (at a constant energy) by 20%.

This design can achieve 289GFLOPS/W at 79 GFLOPS/mm² in low energy mode, and 278GFLOPS/mm² at 60GFLOPS/W in high performance mode. The measurement of the DP throughput unit, DP FMA, is also shown in Fig. 3. Using $V_{DD}$ and BB it achieves 117GFLOPS/W at 13GFLOPS/mm² in low energy mode, and 111GFLOPS/mm² at 20GFLOPS/W.

Fig. 4 provides energy vs. performance tradeoffs for the latency optimized designs. The performance metric is the average delay, which is the product of the clock period and the average cycles per FLOP (i.e. one plus the average latency penalty) when running the SPEC FP benchmarks. The BB control provides energy reduction in two ways. First, it reduces the power by approximately 13% if the unit is heavily used, by lowering $V_{DD}$ and $V_t$. The problem with this statically set BB is shown in the 10% utilization curves. Many applications use FP, but do not use it extensively. In this case, using the same $V_{DD}$ and $V_t$ as the 100% activity core makes the leakage energy dominate, and increases the energy/op by 3x. By dynamically adjusting $V_t$ through the BB (i.e. lowering BB for low-utilization period), one can bring this energy down to 1.5x of the full utilized case.

The measurement results were obtained using a built-in test capability as shown in Fig. 5. High speed on-chip RAMs are implemented to feed/store the inputs/outputs of the selected FPU during a test run (at full FPU speed). A JTAG interface is use to load and check values in the RAMs in a lower speed.

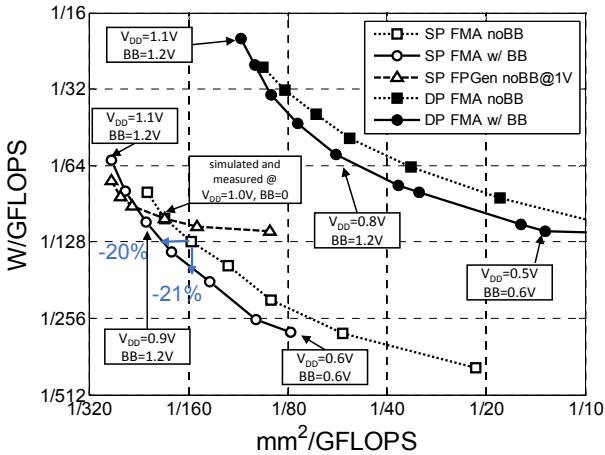

Fig. 3. Throughput tradeoffs for SP and DP FMAs, and the nominal and the peek area/energy efficient operating points.

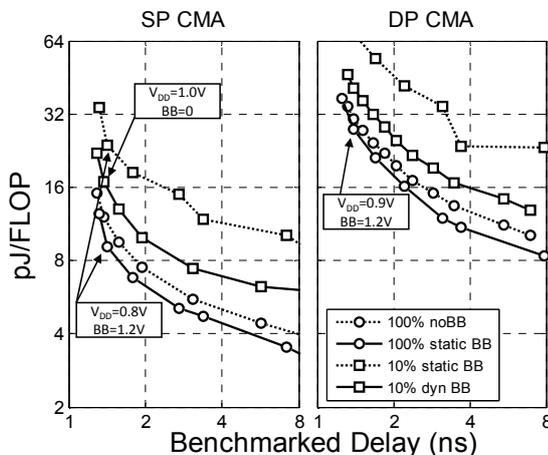

Fig. 4. Latency tradeoffs for SP and DP CMAs at 100% utilization with and without BB, and 10% utilization with statically set BB and dynamically adaptive BB.

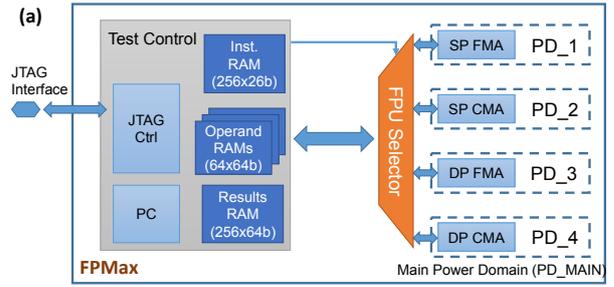

Fig. 5. (a) FPMax chip block diagram, and (b) instruction encoding.

## Conclusion

Table I summarizes the performance of FPMax, and Table II compares the nominal performance of our SP throughput design to other published designs [4-7]. In the comparison, we scale their area and power with the feature sizes and the performance according to FO4, which we expect to provide numbers better than actual silicon. These results demonstrate that our FPU generator creates working designs which match its performance estimates, allowing designers to quickly create designs optimized for their application. It also demonstrates the advantages of strong $V_t$ control, since this allows one to save 20% of the energy when compute bound, but not lose 3x in leakage when the unit is marginally utilized.

## Acknowledgements

This work was supported in part by C-FAR, one of the six STARnet Centers, sponsored by MARCO and DARPA, and by the DARPA SEEC project. The authors would also like to thank ST Microelectronics for fabrication of this chip.

TABLE I. PERFORMANCE SUMMARY

| FPU | DP CMA | DP FMA | SP CMA | SP FMA |
|---|---|---|---|---|
| Area (mm²) | 0.032 | 0.024 | 0.018 | 0.0081 |
| Pipeline Stages | 5 | 6 | 6 | 4 |
| Multiplier Pipe Depth | 2 | 2 | 3 | 2 |
| Adder Pipe Depth | 2 | N/A | 2 | N/A |
| Booth Encoding | 3 | 3 | 2 | 3 |
| Reduction Tree | Wallace | Array | Wallace | ZM |
| Supply Voltage | 0.9V | 0.8V | 0.8V | 0.9V |
| Body-bias | 1.2V | 1.2V | 1.2V | 1.2V |
| Frequency | 1.19GHz | 910MHz | 1.36GHz | 910MHz |
| Leakage Power | 8.4mW | 3.8mW | 3.3mW | 1.6mW |
| Total Power | 66mW | 41mW | 25mW | 17mW |
| **Max**/Norm Area Efficiency (GFLOPS/mm²) | **87.5**/74.6 | **111**/74.6 | **165**/151 | **278**/217 |
| **Max**/Norm Energy Efficiency (GFLOPS/W) | **128**/36.0 | **117**/43.7 | **314**/110 | **289**/106 |
| **Min**/Norm Benchmarked Delay (ns) | **1.18**/1.39 | **1.88**/2.79 | **1.30**/1.42 | **1.39**/1.77 |

TABLE II. PERFORMANCE COMPARISON

| FPU Design | Area Efficiency (GFLOPS/mm²) | Energy Efficiency (GFLOPS/W) |
|---|---|---|
| SP FMA (FPMax) | 217 | 106 |
| Variable-precision FMA [4] | 62.5 | 52.8 |
| Resonant FMA [5] | 142 | 54.9 |
| CELL FMA [6] | 384 | 66 |
| Reconfig. FPU [7] | 98 | 33.7 |